\documentclass[preprint,journal]{vgtc}       

\ifpdf
  \pdfoutput=1\relax                   
  \usepackage{graphicx}                
  \DeclareGraphicsExtensions{.pdf,.png,.jpg,.jpeg} 
\else
  \usepackage{graphicx}                
  \DeclareGraphicsExtensions{.eps}     
\fi%

\graphicspath{{Figures/}{pictures/}{images/}{./}} 

\usepackage{microtype}                 
\PassOptionsToPackage{warn}{textcomp}  
\usepackage{textcomp}                  
\usepackage{mathptmx}                  
\usepackage{times}                     
\usepackage{cite}                      
\usepackage{tabu}                      
\usepackage{booktabs}                  

\usepackage{multirow}
\usepackage{makecell}
\usepackage{enumitem}
\usepackage{dblfloatfix}
\usepackage{placeins}
\usepackage{soul}

\preprinttext{To appear in IEEE Transactions on Visualization and Computer Graphics.}

\onlineid{0}

\vgtccategory{Research}
\vgtcpapertype{application/design study}

\newcommand{\Sibyl}{\textsc{Sibyl}}

\newcolumntype{x}[1]{%
>{\raggedleft\hspace{0pt}}p{#1}}%

\newcolumntype{P}[1]{>{\centering\arraybackslash}p{#1}}

\newcommand{\problem}{usability challenge}
\newcommand{\problems}{usability challenges}
\newcommand{\problemwithoutusability}{challenges}

\newcommand{\page}{interface}
\newcommand{\pages}{interfaces}

\newcommand{\feature}{factor}
\newcommand{\features}{factors}
\newcommand{\featurecap}{Factor}

\title{Sibyl: Understanding and Addressing the Usability Challenges of Machine Learning In High-Stakes Decision Making}

\author{Alexandra Zytek, Dongyu Liu, Rhema Vaithianathan, and Kalyan Veeramachaneni}
\authorfooter{
\item
 Alexandra Zytek, Dongyu Liu, and Kalyan Veeramachaneni are with MIT. E-mail: {zyteka, dongyu}@mit.edu, kalyan@csail.mit.edu.
\item
 Rhema Vaithianathan is with Auckland University of Technology. E-mail: rhema.vaithianathan@aut.ac.nz.
}

\shortauthortitle{Zytek \MakeLowercase{\textit{et al.}}: Understanding Usability Challenges}

\abstract{Machine learning (ML) is being applied to a diverse and ever-growing set of domains. In many cases, domain experts --- who often have no expertise in ML or data science --- are asked to use ML predictions to make high-stakes decisions. Multiple ML usability challenges can appear as result, such as lack of user trust in the model, inability to reconcile human-ML disagreement, and ethical concerns about oversimplification of complex problems to a single algorithm output. In this paper, we investigate the ML usability challenges that present in the domain of child welfare screening through a series of collaborations with child welfare screeners. Following the iterative design process between the ML scientists, visualization researchers, and domain experts (child screeners), we first identified four key ML challenges and honed in on one promising explainable ML technique to address them (local factor contributions). Then we implemented and evaluated our visual analytics tool, \Sibyl{}, to increase the interpretability and interactivity of local factor contributions. The effectiveness of our tool is demonstrated by two formal user studies with 12 non-expert participants and 13 expert participants respectively. Valuable feedback was collected, from which we composed a list of design implications as a useful guideline for researchers who aim to develop an interpretable and interactive visualization tool for ML prediction models deployed for child welfare screeners and other similar domain experts.
} 


\CCScatlist{ 
 \CCScat{K.6.1}{Management of Computing and Information Systems}%
{Project and People Management}{Life Cycle};
 \CCScat{K.7.m}{The Computing Profession}{Miscellaneous}{Ethics}
}

\vgtcinsertpkg

\begin{document}


\maketitle

\definecolor{red}{RGB}{138, 15, 17}
\definecolor{lightred}{RGB}{250, 209, 210}
\definecolor{blue}{RGB}{35, 81, 118}
\definecolor{lightblue}{RGB}{216, 231, 243}
\definecolor{green}{RGB}{47, 108, 45}
\definecolor{lightgreen}{RGB}{220, 240, 219}
\definecolor{purple}{RGB}{96, 50, 103}
\definecolor{lightpurple}{RGB}{236, 221, 238}
\definecolor{orange}{RGB}{153, 76, 0}
\definecolor{lightorange}{RGB}{255, 230, 204}
\definecolor{yellow}{RGB}{140, 140, 13}
\definecolor{lightyellow}{RGB}{251, 251, 208}
\definecolor{brown}{RGB}{123, 64, 30}
\definecolor{lightbrown}{RGB}{245, 225, 214}

\definecolor{highlight}{RGB}{0, 0, 255}
\newcommand{\add}[1]{#1}

\renewcommand{\quote}{\list{}{\rightmargin=\leftmargin\topsep=.3pt}\item\relax}

\newcommand{\TR}{\sethlcolor{lightred}\textcolor{red}{\hl{TR}}}
\newcommand{\DS}{\sethlcolor{lightblue}\textcolor{blue}{\hl{DIS}}}
\newcommand{\CN}{\sethlcolor{lightyellow}\textcolor{yellow}{\hl{CON}}}
\newcommand{\AC}{\sethlcolor{lightorange}\textcolor{orange}{\hl{ACC}}}
\newcommand{\CT}{\sethlcolor{lightpurple}\textcolor{purple}{\hl{CT}}}
\newcommand{\UT}{\sethlcolor{lightgreen}\textcolor{green}{\hl{UT}}}
\newcommand{\ET}{\sethlcolor{lightbrown}\textcolor{brown}{\hl{ETH}}}

\begin{table*}[t!]
\caption{List of \problemwithoutusability{} that could negatively impact the usability of an ML model.}
\begingroup
\renewcommand{\arraystretch}{1.2} 
\begin{tabular}{{x{0.4\textwidth}p{0.05\textwidth}p{0.40\textwidth}}}

\toprule
\textbf{Usability Challenge}                                                                                    & \textbf{Code} & \textbf{Mitigating Tools}                                                                        \\ \midrule
Lack of \textbf{TR}ust                                                                                                 & \TR           & Global explanations, local explanations, performance metrics, historical predictions and results \\
Difficulty \textbf{R}econciling human-ML \textbf{D}isagreements                                                            & \DS           & Local explanations                                                                               \\
\textbf{U}nclear \textbf{C}onsequences of actions                                                                               & \CN           & Cost-benefit analysis, historical predictions and results                                        \\
Lack of \textbf{AC}countability or protections from accountability                                                     & \AC           & Local explanations, performance metrics  
\\
\textbf{E}thical \textbf{C}oncerns (ex. possible bias, concerns about oversimplification)                                       & \ET           & Global explanations, local explanations, ML fairness metrics, historical predictions and results
\\
\textbf{C}onfusing or unclear prediction \textbf{T}arget (ie. the measure predicted by the model has an unclear meaning or significance)   & \CT           & Cost-benefit analysis, further analysis of prediction target impact                              \\
\textbf{U}nhelpful prediction \textbf{T}arget (ie. the measure predicted by the model is not relevant to the required decision) & \UT           & Retrain model with new prediction target                                                          \\ \bottomrule
\end{tabular}
\endgroup
\label{tab:usability_concerns}
\end{table*}


\section{Introduction}
Thanks to innovations in machine learning (ML), computers can now help with many tasks previously performed by humans alone, often improving both speed and precision. However, in many domains, human decision-makers provide essential insights that cannot be replaced by existing algorithms. In such cases, decision-making outcomes are improved when ML output is used to augment human decision-making, rather than replace it. 

ML models often output only a single number or classification, such as the risk score seen in the upper right corner of Figure \ref{fig:teaser}. This can make it difficult for human decision-makers to incorporate the model into their decision making. As a result, many ML algorithms lack \textit{usability}, or the attribute of being able to be efficiently used by humans to make better decisions. 

The machine learning, data science, and data visualization communities have offered a multitude of algorithms and tools to augment ML predictions and address these \problems{} --- we refer to these as \textbf{ML augmentation tools}. These tools, when chosen carefully for the domain, have the ability to greatly improve the usability of ML models for decision making. Examples of such tools include data visualizations, global and local explanations \cite{doshi-velezRigorousScienceInterpretable2017}, cost-benefit analysis \cite{kentonHowCostBenefitAnalysis2021}, performance metrics, and information about historic usage and results of the ML model. However, research aimed at augmenting ML predictions often focuses on an audience of ML/data experts \cite{zhangManifoldModelAgnosticFramework2019} \cite{strobeltLSTMVisToolVisual2017} \cite{kahngActiVisVisualExploration2017} or domain experts in more technical or data-driven fields such as medicine \cite{lundbergExplainableMachinelearningPredictions2018} \cite{kwonRetainVisVisualAnalytics2019}. For example, Zhang et. al. \cite{zhangManifoldModelAgnosticFramework2019} developed a framework for helping data scientists and ML experts interpret and debug ML models, and Lundberg et. al. \cite{lundbergExplainableMachinelearningPredictions2018} developed an interface for helping anaesthesiologists prevent hypoxaemia during surgery through detailed data visualization. In contrast, many fields are more qualitative in nature, with decisions following discussion more than data crunching. In this paper, we focus on these more qualitative fields, and the \problems{} they face.

\add{To concretely assess usability challenges, we investigated the need for additional auxiliary information alongside ML predictions through a comprehensive literature review. We first selected a number of papers that had ML and explainability as topics. We selected and reviewed 55 papers covering ML applications and explainability. We identified challenges in human-ML interactions described in these papers. Three of us then began codifying a set of factors that decrease usability in ML models. Table \ref{tab:usability_concerns} summarizes a set of \problems{} we codified that are relevant when a model is actively used for decision-making. }

Some \problemwithoutusability{} stem from not understanding where a model's predictions come from, making it difficult for human decision-makers to trust the model (\TR{}), and to handle any disagreements between their opinions and the model's output (\DS{}). Others are caused by a lack of information about the real effects of a decision. A lone model prediction often does not explicitly indicate the expected results of a decision (\CN{}), suggest accountability (\AC{}), or provide ethical assurances (\ET{}). Finally, challenges may arise when the output of the model is not a direct suggestion of a decision, but rather auxiliary information. In this case, the output may be confusing (\CT{}) or entirely irrelevant (\UT{}). 


Determining which \problems{} exist, the best tools to address them, and the necessary design choices for these tools depends highly on specific aspects of the domain and the decision-makers involved. \add{Through our literature review and the case study discussed in this paper, we identified a subset of context-dependent factors that should be considered when working to make an ML model more usable in a particular domain. Table \ref{tab:context_factors_table} lists some examples of these factors. }

To investigate the problem of finding and mitigating \problems{} in more qualitative fields, we selected the domain of child welfare screening. In terms of the relevant context factors, child welfare screeners are domain experts without ML/data science expertise, making decisions using an ML model as an auxiliary tool, with about a few minutes per decision, in a high-risk field. 

Addressing \problems{} is a non-trivial task that requires collaboration with end-users. In this paper, we engaged in three forms of collaboration: \textbf{observations} to understand their existing workflow and its possible \problems{}, \textbf{interviews} to gain additional insights into the desires of end-users, and \textbf{user studies} with possible ML augmentation techniques to get concrete feedback on design. 

\textbf{The main research questions, and our key findings with regards to these questions, are as follows. }
\begin{enumerate}[label=\textbf{RQ\arabic*}]
    \item What ML \problems{} exist in the domain of child welfare screening? 
    Through interviews and field observations, our work identifies four main challenges, described in Section \ref{sec:context}. 
    \item What interfaces can be helpful in mitigating these ML \problems{}? We designed and implemented five augmentation interfaces, making up our \Sibyl{} tool, described in Section \ref{sec:sibyl}. Based on our interviews and a formal user study, described in Section \ref{sec:user_study_1}, we suggest that local \feature{} contributions are most useful for this domain.
    \item What design choices must be made when building these tools to optimize them for use by child welfare screeners and other experts in similar domains? Based on our interviews and formal user study, we identified a list of design considerations that should be made when choosing model features and developing ML augmentation visualizations, described in Sections \ref{sec:user_study_1} and \ref{sec:discussion}.
\end{enumerate}

\begin{table*}[t]
\caption{Domain context factors that may influence the usability of an ML model. The context factors relevant to child welfare screening are in bold. By \textit{technical expertise}, we refer to the ML or data science expertise of the end-user. This is not meant to be a complete list --- there are many other factors that could also be relevant. }
\footnotesize
\begin{tabular}{{x{0.18\textwidth}p{0.45\textwidth}p{0.3\textwidth}}}

\toprule
\textbf{Context Factors}              & \textbf{Categorizations}                                               & \textbf{Example domain}           \\ \midrule
\multirow{5}{*}{Degree of Automation} & Fully autonomous, humans not involved in decision making                & Level 5 self-driving car \cite{nationalhighwaytrafficsafetyadministrationnhtsaAutomatedVehiclesSafety2017}                              \\
                                      & Machine makes decisions, humans monitor                                & Level 4 self-driving car \cite{nationalhighwaytrafficsafetyadministrationnhtsaAutomatedVehiclesSafety2017}            \\
                                      & Machine suggests decisions, humans make decisions                      & Content violation flagging                 \\
                                      & \textbf{Machine provides auxiliary information, humans make decisions} & Child welfare screening                  \\ \midrule
\multirow{4}{*}{Decision time}        & Immediate                                                              & Level 5 self-driving car \cite{nationalhighwaytrafficsafetyadministrationnhtsaAutomatedVehiclesSafety2017}                                     \\
                                      & Seconds                                                                & Aircraft emergency response                         \\
                                      & \textbf{Minutes}                                                       & Child welfare screening                               \\
                                      & Hours                                                                  & Non-emergency medical procedures                     \\ \midrule
\multirow{3}{*}{\makecell[r]{Technical expertise \\of humans}} & \textbf{Little to none}         & Child welfare screening                                                  \\
                                      & Experience with data science                                           & Finance                                                           \\
                                      & Machine learning/Data science expert                                   & ML model training and debugging                                           \\ \midrule
\multirow{3}{*}{\makecell[r]{Domain expertise \\of humans}}                & Little to none                       & Autonomous aircraft (to passengers)                                         \\
                                      & Basic understanding/Intuition                                          & Crowdsourcing                                               \\
                                      & \textbf{Domain expert}                                                 & Child welfare screening                             \\ \midrule
\multirow{3}{*}{Associated Risk}                       & Low                                                                    & Camera roll image sorting                                       \\
                                      & Medium                                                                 & Mail sorting                                                       \\
                                      & \textbf{High}                                                                   & Emergency medical procedures                                     \\ \bottomrule
\end{tabular}
\label{tab:context_factors_table}
\end{table*}

\section{Study Context: Child Welfare Screening} \label{sec:welfare}
In this section, we introduce the domain of child welfare screening

\textbf{Child Abuse in the U.S.} Child abuse is an active issue affecting the health and well-being of communities. The Centers for Disease Control and Prevention (CDC) estimates at least 1 in 7 children have experienced child abuse and/or neglect in the past year \cite{childrensbureauChildMaltreatment20182020}. Child abuse victims can suffer physical and emotional injuries, and may experience trauma resulting in long-term mental health problems \cite{childrensbureauChildMaltreatment20182020}. More than one-third of American children are investigated as potential victims of abuse or neglect by age 18 \cite{kimLifetimePrevalenceInvestigating2017}. Still, in 2018, there were 1,770 reported fatalities resulting from child abuse and neglect \cite{childrensbureauChildMaltreatment20182020}. 

\textbf{Child Welfare Screening.} In the U.S., regional Child Protective Services (CPS) agencies are tasked with handling child abuse and neglect referrals from concerned members of the community, including mandated reporters such as teachers, who are required by law to report any suspicion of abuse or neglect. These referrals are examined by child welfare specialists (``call screeners"), who decide whether to screen in or screen out each case. A screened-in case will be investigated further, while a screened-out case will be recorded but not investigated. 

Both false negatives (real abuse cases that are screened out) and false positives (cases with no abuse that are screened in) can have heavy consequences. False negatives lead to prolonged child suffering and, in extreme cases, child fatality. False positives can lead to long-term emotional distress for parents, children, and other family and community members, as well as damaged financial, career and social prospects for parents and other caretakers \cite{richardsonEffectsFalseAllegation1990}.

In 2018, CPS agencies in the United States received 4.3 million referrals from concerned parties about potential child abuse \cite{childrensbureauChildMaltreatment20182020}. 56\% of these referrals were screened in and investigated, but only 16.8\% of the screened in cases were found to involve abuse or neglect \cite{childrensbureauChildMaltreatment20182020}. 

\textbf{ML for Child Welfare}. 
One important motivation for computerized assistance in child welfare call screening is repeated cases of missed abuse. Fatal child abuse cases in which children were referred several times but were never screened in are tragic, and although such cases are rare, they are avoidable \cite{hurleyCanAlgorithmTell2018}. An ML solution can quickly scan for red flags, such as repeated referrals, that busy human call screeners may miss in the overload of data.

In recent years, predictive risk modelling (PRM) has been deployed in child welfare contexts in multiple counties, with the goal of enabling more efficient and consistent decision-making and improving the overall health and safety of county residents \cite{vaithianathanDevelopingPredictiveModels2017}. One example of such a model was deployed in Allegheny Country, PA by Vaithianathan et. al. in 2016 \cite{vaithianathanDevelopingPredictiveModels2017}.

Currently, PRM is being introduced to our collaborating county in Colorado by Vaithianathan et. al. \cite{vaithianathanImplementingChildWelfare2019}, through a LASSO regression model trained on 461 features, which include information such as the child and parents' age, past referrals and their outcome, and past court involvements. The model predicts the likelihood of removal from home in the next two years, translated to a 1 through 20 risk score where the higher the score, the higher the risk \cite{vaithianathanDevelopingPredictiveModels2017}. This paper focuses on the usage of this model.

\textbf{Study Participants.}
We collaborated over the course of a year with a pool of 19 social workers and supervisors working for the child welfare department in a collaborating county in Colorado. All participants regularly act as screeners in the county's child welfare screening decision making process. 
Our collaborations began in December 2019, with two days of in-person field observations (Section \ref{sec:observation}). Following this, we observed a simulated case review session via video conferencing (Section \ref{sec:usability}), and conducted several interviews also via video conferencing (Sections \ref{sec:interviews} and \ref{sec:sibyl}). Finally, our collaborating screeners participated in our user study digitally (Section \ref{sec:user_study_1}).

\section{Related Work} \label{sec:related_work}
In this section, we discuss related work in human-centered and explainable ML, as well as existing ML augmentation tools.

\textbf{Human-Centered ML.}
Past literature has advocated for a \textit{human-centered} perspective to ML \cite{gilliesHumanCentredMachineLearning2016} --- one that considers machines and algorithms as part of collaborative systems alongside humans. This  perspective considers how humans use, interact with, adapt to, and evaluate ML applications \cite{gilliesHumanCentredMachineLearning2016}. A truly human-centered ML approach acts end-to-end, beginning with human-in-the-loop training systems and ending with evaluation systems based on the metrics that end users are most interested in \cite{gilliesHumanCentredMachineLearning2016}. In this paper, we take a deeper look at one step of this extensive pipeline: the use of ML algorithm predictions by humans for real-world decision making. 

\textbf{Explainable ML.} A common \problem{} addressed by the literature is the black box nature of most ML algorithms. Humans struggle to use ML predictions because they do not understand where they came from. This \problem{} is addressed through the fields of interpretable or explainable ML. Doshi-Velez and Kim proposed that the need for ML interpretability stems from an ``incompleteness in the problem formulation'' \cite{doshi-velezRigorousScienceInterpretable2017}, which prevents the system from being thoroughly evaluated or trusted. This incompleteness can take several forms, including a need for scientific understanding, concerns about safety or ethics, or mismatched objectives between the model output and the human goal \cite{doshi-velezRigorousScienceInterpretable2017}. 

Doshi-Velez and Kim \cite{doshi-velezRigorousScienceInterpretable2017} also define three evaluation approaches for ML interpretability. In \textit{application-grounded approaches}, domain experts work with explanations within a real application. This provides the most realistic quantification of explanation quality, but may require high time commitments from a potentially small pool of domain experts. In \textit{human-grounded approaches}, researchers develop simpler problems for experimentation using non-expert subjects. Finally, in \textit{functionally-grounded evaluation}, a formal definition of interpretability is used as a proxy to evaluate an explanation without using human subjects. This paper focuses on an example of such an application-grounded approach. 

Wang et. al. \cite{wangDesigningTheoryDrivenUserCentric2019} developed a human-driven conceptual framework for building explainable AI systems. They found that decision-makers seek explanations to justify unexpected occurrences, monitor for important events, or facilitate learning. They created a taxonomy of AI techniques based on how they support human reasoning and represent information. 
Finally, the authors discuss how explainable AI can mitigate cognitive biases. Our work builds on this by finding cognitive biases that can be \textit{caused} by explainable AI, as listed in Section \ref{sec:discussion}. 

\textbf{ML Augmentation Tools. }
\add{Spinner et. al. \cite{spinnerExplAInerVisualAnalytics2020} developed a conceptual framework for explainable AI, along with a corresponding implementation called \textsc{explAIner}. This framework includes single-model explainers (which are the focus of this paper) as well as multi-model explainers, which can be used to compare and select between models. The authors also distinguish between different model audiences, including novices, model users, and developers --- our work focuses on model users.}

\add{Krause et. al. \cite{krauseInteractingPredictionsVisual2016} developed \textsc{Prospector}, a comprehensive visualization system for data scientists.  This system includes interactive functionality for both global and local feature-focused explanations.}

Hohman et. al. \cite{hohmanGamutDesignProbe2019} developed a visual analytics system called \textsc{Gamut} to investigate how machine learning practitioners and data scientists interact with machine learning. To develop this tool for use on GAMs, the authors interviewed technical experts to generate a list of common questions asked about predictions. In total, they identified six question types, which they address using three views. 
\textsc{Gamut} was tested by having 12 data scientists use the tool while thinking out loud, followed by an interview. 

Lundberg et. al. developed \textsc{Prescience}, an explanatory ML system focused on preventing hypoxaemia during surgeries \cite{lundbergExplainableMachinelearningPredictions2018}. This tool predicts the risk of hypoxaemia in the next five minutes using a gradient-boosting algorithm trained on time series. It also includes several visualizations to explain the prediction, including SHAP feature contribution explanations.

Kwon et. al. \cite{kwonRetainVisVisualAnalytics2019} developed \textsc{RetainVis}, a visualization tool for explaining recurrent neural networks (RNNs) applied to electronic medical records (EMRs). The tool was developed with active feedback from domain experts (medical practioners). \textsc{RetainVis} includes five different visualizations for looking into RNNs.

\add{Our work is similar to these tools in that it relies on collaboration with end users to develop a tool that provides additional information alongside an ML prediction. However, our users are not expected to have any prior ML or data science expertise, nor are they used to working with data-heavy visualizations. This work is also the first to our knowledge that investigates through a complete case study the usability of machine learning for child welfare screeners.}
\section{Understanding Context and End-User Needs}\label{sec:context}

To address \textbf{RQ1} (identifying ML \problems{}), we performed a series of field observations and interviews. In this section, we discuss our goals and the findings we made during these steps.

\begin{figure*}[t]
    \centering
    \vspace{-0.2cm}
    \includegraphics[width=0.9\linewidth]{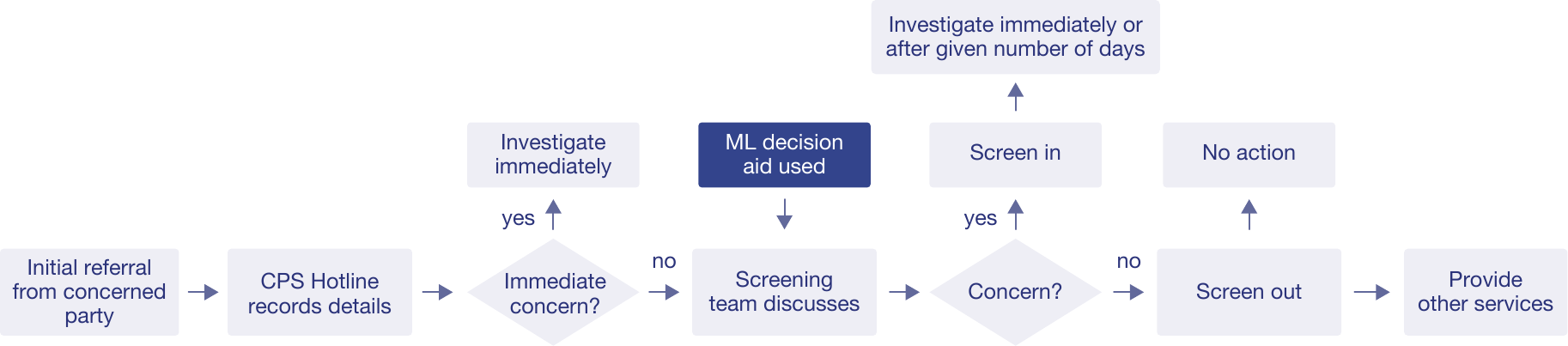}
    \caption{The general child welfare screening process used by the CPS department of our collaborating county. The referral is first received by the CPS hotline, and then sent to a child welfare supervisor. In a minority of cases, the supervisor will deem the child or children involved to be in serious and immediate risk of danger, and will screen-in the case for immediate further investigation. In most cases, however, the case will be reviewed by a team of child welfare screeners the next day. This team will be given the ML risk score prediction (dark blue box). If this team decides to screen-in the case, it will be investigated further through home visits, interviews, or other means. Otherwise, the case will be recorded but not investigated unless re-referred. In the case of a screen-out, the screeners may elect to provide the family with additional family services.} 
    \label{fig:screening_process}
    \vspace{-0.5cm}
\end{figure*}

\subsection{Understanding Existing Workflows} \label{sec:observation}

To better understand the existing child welfare screening workflow, we travelled to our collaborating county in Colorado to observe screeners' decision-making on referrals without using the ML model. This process led to the following findings:

\begin{enumerate}
    \item Our collaborating county uses the general procedure for child welfare screening shown in Figure \ref{fig:screening_process}. In cases of immediate concern, a referral may be screened in immediately after it is received by CPS (this decision is made by a child welfare supervisor). In most cases, however, the decision as to whether to screen-in or screen-out a referral is made by a team of child welfare experts. It is this team that receives the ML risk score prediction.
    \item Five to ten minutes are spent on each case by this team. Most of this time is spent going over the details of the case. The screening decision is made after about one to two minutes of discussion. 
    \item  A large portion of these five-to-ten minutes of screening time is dedicated to determining the factors that are associated with higher and lower likelihood of abuse --- referred to as risk and protective factors, respectively --- involved in a case, and weighing these against each other. The factors considered will vary based on the details of the case, but may include information such as child's age (very young children are more vulnerable), criminal record of adults involved, whether there are any trusted adults active in the child's life, and actions and statements made by adults involved (for example, a mother taking actions to separate an aggressive partner from her children).
\end{enumerate}

\subsection{Identifying ML Usability Challenges} \label{sec:usability}
To identify the particular ML \problems{} relevant to child welfare screening (\textbf{RQ1}), we observed a simulated case review session where social workers used an ML model.  \add{In this session, our 19 expert collaborators were split into three teams of typical size, which we will refer to as \textbf{T1}, \textbf{T2}, and \textbf{T3}. They were asked to make decisions about real referrals that the county had handled in the past. }
\add{For each case, they had access to the information that screeners are typically given during decision-making — including the age of all involved parties and a written description of the potential abuse as given by the referring party — as well as a 1 through 20 risk score provided by the ML model. After receiving this information, the screeners went ahead with their usual process: discussing the case as a team for five to ten minutes and then making a screening decision. The teams were then interviewed and asked to reflect upon how they used each ML score, whether the scores aligned with their expectations, and how the scores impacted their decisions. Each team made decisions on seven to nine cases. }

\add{During interview sessions, all three teams expressed reservations about using the ML model for decision-making. Based on their responses to our interview questions, we identified four key usability challenges:}

\begin{enumerate}
    \item \textbf{Lack of Trust \TR \quad} Screeners expressed a lack of trust when making decisions using the ML model, evidenced by their tendency to not consider the model prediction at all when it disagreed with their intuition. For example, when asked if the score caused them to reconsider their decision, \textbf{T3} responded
    \begin{quote}
        \textit{``No, [we were] surprised it is that low.''}
    \end{quote} 

    \item \textbf{Difficulty reconciling human-ML disagreements \DS \quad} Screeners did not have a clear path forward when they disagreed with the model prediction, sometimes electing to ignore it entirely (see previous item) and sometimes trying to justify it based on how they thought the model worked. For example, \textbf{T2} reported in one case that the score made them
    \begin{quote}
        \textit{``think a little deeper about why the score is so high [and caused us to] take another look at [the history]''}
    \end{quote}
    
    \item \textbf{Unclear prediction target \CT \quad} Because the model provides auxiliary information (1-20 risk score based on the likelihood of removal from home in 2 years) rather than a direct decision suggestion, there was some confusion about how to use the model prediction target. For example, when asked how the model affected their decision making process, \textbf{T3} responded 
    \begin{quote}
        \textit{``[we did not know] enough of what the score means to know how to accurately use it.''}
    \end{quote}
    \textbf{T3} also said they
    \begin{quote}
        \textit{``Wish we knew how we got to the score.''}
    \end{quote}
    \item \textbf{Concerns about Ethics \ET \quad} As expected for such a sensitive domain, users were concerned about the ethics of using the ML model score. There was concern that the model may prevent critical thinking. \textbf{T3} commented
    \begin{quote}
        \textit{``[The model] could be dangerous for people just looking at the number, need to take everything into account. Makes you stop and think and ask yourself are you critically thinking...''}
    \end{quote}
\end{enumerate}

\add{In addition to determining which usability challenges were relevant to child welfare screening, we also confirmed that some were \textit{not} relevant, and therefore did not require consideration when designing \Sibyl{}. For example, three of the example usability challenges from Table \ref{tab:usability_concerns} were not relevant. Lack of accountability \AC{} was not relevant, as accountability is always held by human decision-makers in this domain. Similarly, unclear consequences of actions \CN{} was not relevant, because the domain experts have extensive training and understanding of the potential consequences of screening decisions, and do not intend to offload this understanding to an ML algorithm. Finally, unhelpful prediction target \UT{} was not relevant, as the ML model has been developed specifically to provide relevant information.}

\begin{figure*}[!htb]
    \centering    
    \vspace{-0.2cm}
    \includegraphics[width=0.75\linewidth]{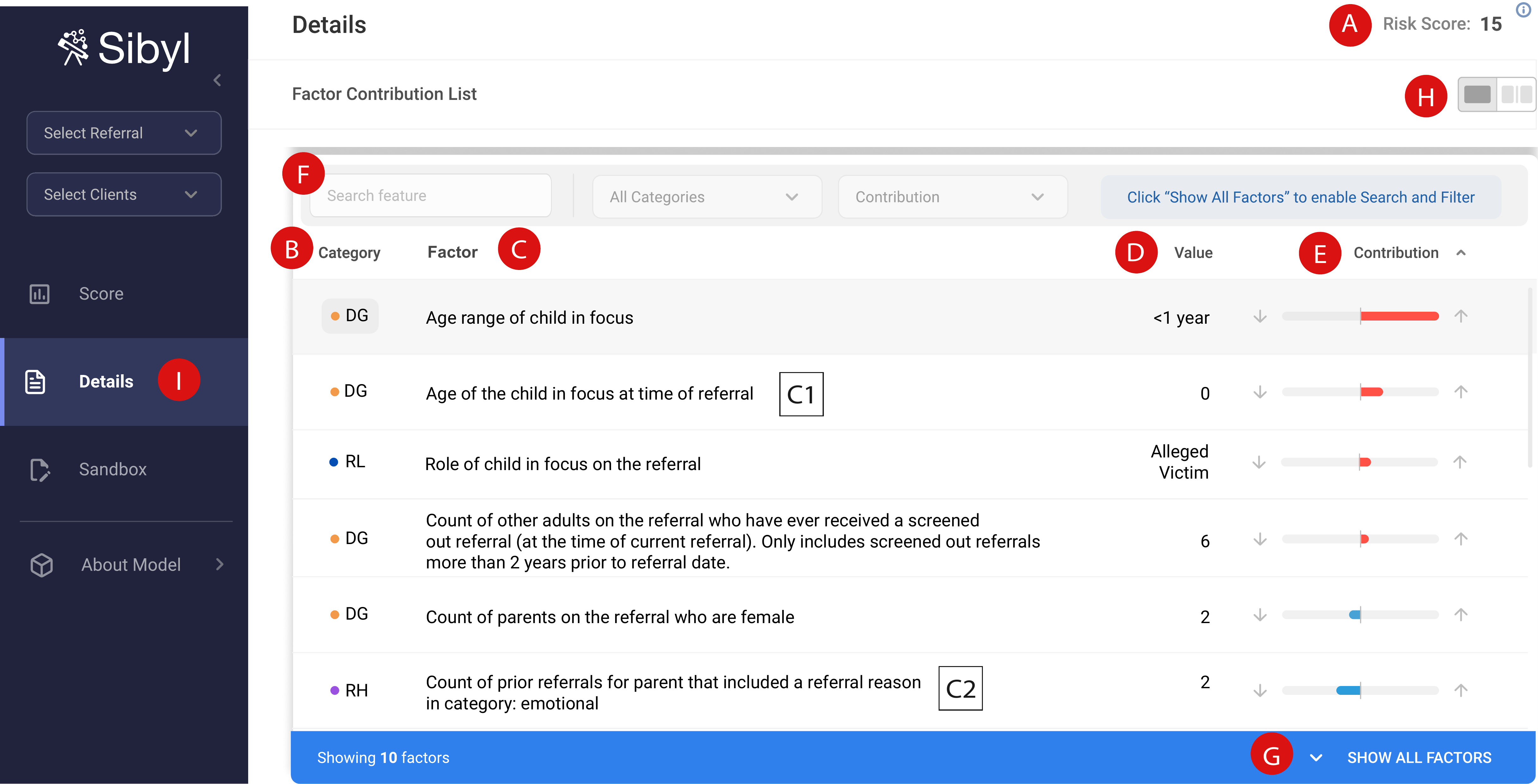}
    \caption{\add{The final version of the "Case-Specific Details" interface after factoring in user study feedback.} This interface shows how particular factors contribute to predictions made by ML models about child welfare. Labeled elements are as follows: (A) The risk score for the case (1-20). (B) Categories for each factor, such as demographics (DG) or referral history. (C) A short description of each factor. (D) The value of numeric or categorical factors. (E) The contribution of each factor (the table can be sorted in ascending or descending order of contribution). (F) UI components for searching by factor name or filtering by category, enabled when the full factor list is shown. (G) A button for switching between a view that shows only the top 10 most contributing factors and one that shows all factors. (H) A button for switching between a single-table view and a side-by-side view, which splits factors that increase and decrease risk. (I) \add{A sidebar for switching between different explanation types, as described Section~\ref{sec:sibyl}.}}
    \label{fig:teaser}
    \vspace{-0.6cm}
\end{figure*}

\subsection{Interviewing Screeners} \label{sec:interviews}

\add{To begin addressing \textbf{RQ2} (what tools can be helpful in mitigating \problems{}), we interviewed the 19 screeners about what additional information they would be interested in receiving alongside ML predictions. The format was a semi-structured open-floor session. We began by asking the screeners whether they thought additional information would be useful, and what specific information they would be interested in. We also proposed possible augmentation information (e.g. the relative importance of different factors, answers to what-if questions, and comparisons with past cases) and asked if they might be helpful.}

Our findings from this interview included:
\begin{enumerate}
    \item Screeners were confident that they would want to know why the model made the predictions it made.
    \item Screeners believed that understanding how important each \feature{} was to the score prediction would be helpful.
    \item Screeners wanted to know what steps the model takes in making predictions. \label{item:steps}
    \item Screeners were interested in getting ``what-if'' style explanations that give information about what could be changed about a child to reduce his or her risk. Note that the ML model is not trained on causal relationships, so explanations would not be able to provide such information.
    \item Some screeners were interested in seeing similar cases they dealt with in the past. Others thought this would be too much information to digest in such a short period of time.
\end{enumerate}

\section{Getting Feedback on Possible Tools} \label{sec:sibyl}

\begin{table*}[t]
\caption[The proposed \Sibyl{} \pages{}, and the challenges they were theorized to address.]{The proposed \Sibyl{} \pages{} (left column), the challenges they were theorized to address (middle column, using the codes from Table \ref{tab:usability_concerns}), and the reasons we expected these \pages{} to address the given challenges (right column). \TR: Lack of trust in the model. \DS: Difficulty reconciling disagreements. \CT: Confusing prediction target. \ET: Ethical concerns.}
\begin{tabular}{x{0.25\linewidth}P{0.15\linewidth}p{.50\linewidth}}
\toprule
\textbf{Interface}       & \textbf{Challenges Addressed}  & \textbf{How does it address the challenge?}                                             \\
& \TR{} & \\
\midrule
Case-Specific Details    & \TR                            & Reveals relevancy of considered factors                              \\
                         & \DS                            & Highlights factors that may have been missed or misused              \\
                         & \CT                            & Translates score to a concrete factor list                           \\
                         & \ET                            & Allows for critical thinking about factors and score                 \\ \midrule
Sandbox                  & \DS                            & Allows users to test theorized justifications                        \\
                         & \ET                            & Allows for thinking through what-if scenarios                        \\ \midrule
Similar Cases            & \TR                            & Provides information on past performance                             \\
                         & \ET                            & Provides a deeper look into the nuance of cases                      \\ \midrule
Global Factor Importance & \TR                            & Reveals how the model generally makes predictions                    \\
                         & \CT                            & Translates the score to a concrete factor list                       \\ \midrule
Factor Distributions     & \TR                            & Shows how well the model performed on past cases                     \\
                         & \CT                            & Shows the relationship between the risk score and removals from home \\ \bottomrule
\end{tabular}
\label{tab:interfaces}
\end{table*}

To address \textbf{RQ2} (identifying helpful interfaces) and \textbf{RQ3} (identifying important design choices), and based on the \problems{} identified (Section \ref{sec:usability}) and responses to our interview (Section \ref{sec:interviews}), we engaged in a user-centered iterative design process \cite{munznerNestedModelVisualization2009} to develop \Sibyl{}, an ML augmentation tool. We began by designing high-fidelity mock-ups for five augmentation \pages{}, each with a separate purpose and goal. Table \ref{tab:interfaces} summarizes the motivation behind each \page{} in terms of its theorized effect on addressing the \problems{}. The full versions of the original high fidelity mock-ups can be found in Appendix \ref{app:mockups}.


Early in the design process, we learned that the word ``factor'' is more familiar to screeners than ``feature'' when referring to pieces of information used when making decisions. For the purposes of consistency, we use the word factor throughout this paper when referring to data inputs used by the model.

\subsection{Case-Specific Details: \featurecap{} Contributions}
The \textit{Case-Specific Details} \page{} (Figure \ref{fig:teaser}) provides a simple local explanation of where an individual model prediction comes from through \feature{} contributions. 
\add{
The table assigns each \feature{} a contribution (Figure \ref{fig:teaser}E), either positive (red) or negative (blue). The bar length indicates the magnitude of the contribution.
}

\add{The phrases \textit{positive contribution} and \textit{negative contribution} may cause confusion. In the ML community, a positive contribution indicates that the model prediction will increase in value. In the screener community, however, an increased risk score is a negative occurrence. We decided to avoid using the terms ``positive'' and ``negative'' in the app or instructions. Instead, the \features{} are labelled as ``risk factors'' or ``protective factors’’ to mirror the screeners' language. We utilize the bar's direction along with the two arrows ($\uparrow$ and $\downarrow$) to suggest risk increasing (red bar pointing right) or decreasing (blue bar pointing left).}

A local \feature{} contribution explanation may reveal that the young age of a particular child (infant) has caused a significant increase in the risk score (Figure~\ref{fig:teaser}-C1), while the low number of past referrals  (2) compared to the average referred child resulted in a decrease compared to the average risk (Figure~\ref{fig:teaser}-C2).

The local \feature{} contributions were found using the Shapely Additive Explanations (SHAP) algorithm \cite{lundbergUnifiedApproachInterpreting2017}. We chose to use SHAP because it is theoretically grounded in game theory and generates consistent and intuitive explanations for an ML model. 

We decided on a local contribution \page{} as we theorized it may help with all identified \problems{}: the screeners' lack of trust \TR{} by demonstrating that the model relies in part on similar factors as the human screeners in making decisions, difficulty reconciling disagreements \DS{} by highlighting differences in the human and model's logic, unclear prediction target \CT{} by providing a concrete explanation of the scores' meaning, and concerns about ethics \ET{} by making critical thinking about relevant factors easier. 

\add{
\textbf{Feedback.}
Initial interview feedback suggested that screeners were most interested in the Case-Specific Details \page{}, so it was kept as the default option. Additionally, screeners said that in their usual workflow, they would list ``Risk'' and ``Protective'' factors side by side. To mirror this, the updated version of this \page{} has a split-view toggle (Figure~\ref{fig:teaser}H) that shows negatively and positively contributing \features{} in two side-by-side tables). 
}

\subsection{Sandbox: Investigating ``What Ifs''}

\begin{figure}
    \centering    
    \includegraphics[width=1\linewidth]{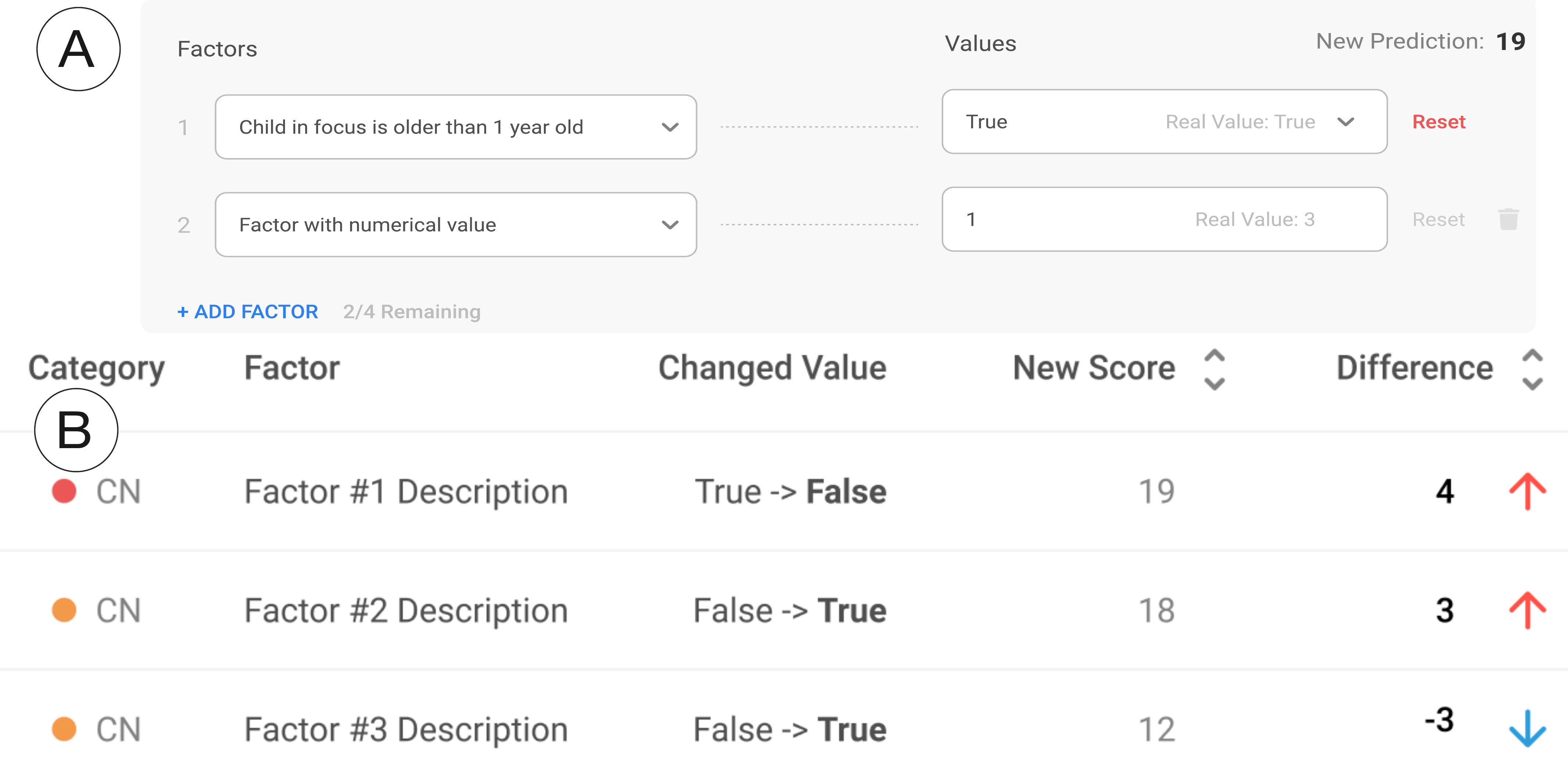}
    \caption{Sandbox visualizations. (A) Users can change up to four factor values at a time, and the new score will be  displayed on the top right corner. (B) The table lists the resulting prediction if each Boolean value is individually reversed.}
    \label{fig:sandbox_visualization}
    \vspace{-0.6cm}
\end{figure}

The \textit{Sandbox} \page{} allows users to experiment with and see how the model prediction would change if \features{} differed. It has two parts: 
\add{
(1) The \textit{Experiment with Changes} box (Figure~\ref{fig:sandbox_visualization}A) allows users to change up to four \feature{} values at a time, to investigate specific ``what-if'' questions.
(2) The \textit{Model predictions if each value was changed} box (Figure~\ref{fig:sandbox_visualization}B) shows the resulting prediction if each Boolean \feature{} value was individually reversed (ie. \texttt{true} to \texttt{false} or \texttt{false} to \texttt{true}).
Following the consistent design principle, we adopted a similar tabular design and use the red/blue up/down arrows to highlight changes in risk scores.
}

This \page{} was added based on feedback from the interviews described in Section \ref{sec:interviews}. We theorized it may help with difficulty reconciling disagreements \DS{} by allowing screeners to test their theorized justifications, and with concerns about ethics \ET{} by making more detailed consideration about the model's output easier.

\add{
\textbf{Feedback.}
Screeners expressed concern that the Sandbox \page{} may be misconstrued as suggesting specific actions or reflecting real-world causal structures. However, screeners also said that they saw value in this \page{} as a supervision tool, used to review model predictions and human decisions rather than being actively used during the decision-making process.
}

\subsection{Similar Cases: Investigating Past Cases} 
The \textit{Similar Cases} \page{} shows the complete history of child welfare involvement with past cases that had similar \feature{} values. 
The similar cases are found using a Nearest Neighbors algorithm. For design purposes, the algorithm used weights all \features{} equally. 
This \page{} includes a timeline for the current case and each similar case, and highlights events such as referrals to child welfare services, investigations, and removals.
\add{
To facilitate comparative analysis, cases are lined up row by row and share the same timeline.
} 

The \page{} was added as we theorized it may help with screeners' lack of trust \TR{} by demonstrating past performance, and concerns about ethics \ET{} by providing a deeper look into an individual case.

\add{
\textbf{Feedback.}
Screeners were concerned that this \page{} may cause poor decision-making. They explained that basing decisions about a current case on past cases that seem similar is discouraged, as this can lead to biases or self-fulfilling prophecies. Therefore, it was decided that this \page{} would not be included in a decision-making tool. However, county officials pointed out that it could be used retroactively (outside of decision-making) to investigate unusual predictions made by the model for the purposes of model evaluation. Due to these concerns, we decided not to include this interface in our formal user study.}

\subsection{Global \featurecap{} Importances: Understanding the Model}
The \textit{About Model} \pages{} offer information about the model's general logic, outside of the context of a individual prediction. 

The first \textit{About Model} \page{} is the \textit{Global \featurecap{} Importance} explanation.
This \page{} shows a global explanation in the form of the general, relative importance of each \feature{}. 
It also provides a brief description of the model architecture and logic, as well as its performance metrics. A visualization for this \page{} can be found in the Appendix, Figure \ref{fig:first_importance}

The global \feature{} importance rankings were found using the Permutation Importance algorithm \cite{fisherAllModelsAre2019}. This algorithm computes the change in model performance if each \feature{} is permuted individually. It therefore describes how closely each \feature{} is linked to model performance.  

This \page{} was added as we theorized it may help screeners build trust in the model \TR{} by seeing how it generally makes predictions, and because it may clarify the meaning of the prediction target \CT{}.

\add{
\textbf{Feedback.}
Screeners said that the Factor Importance \page{} seemed intuitive, but may provide too much information to be practical during active decision-making. Instead, they said it may be useful for training and education. 
}

\subsection{\featurecap{} Distributions: Understanding Past Predictions}
\begin{figure}
    \centering    \includegraphics[width=1\linewidth]{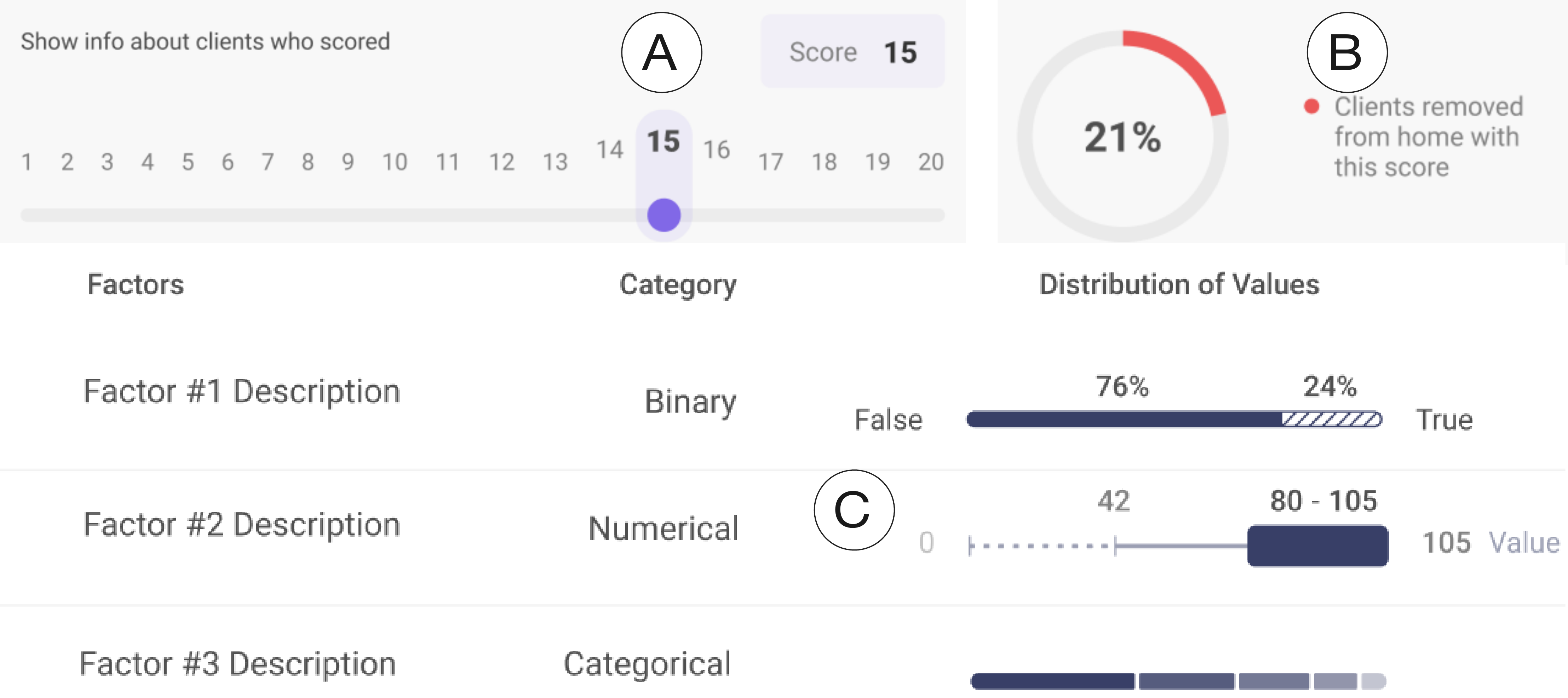}
    \caption{\featurecap{} Distribution visualizations. 
    Users specify a risk score of interest (A) and observe the percentage of children with that score who were removed from the home (B). Three kinds of visualizations (C) are proposed to show the value distribution of children with the selected prediction score.
    }
    \label{fig:feature_distribution_visualizations}
    \vspace{-0.6cm}
\end{figure}

The second \textit{About Model} \page{} is the \textbf{\featurecap{} Distributions} explanation, which gives a quick retrospective view of how the model performed in the past. 
\add{
This \page{} shows the distribution of \feature{} values among past cases that were given a particular score (Figure~\ref{fig:feature_distribution_visualizations}A), as well as the percentage of children with that score who were removed from the home (Figure~\ref{fig:feature_distribution_visualizations}B). 
}

\add{
Depending on the \feature{} type (Figure~\ref{fig:feature_distribution_visualizations}C), the \featurecap{} Distributions explanation uses one of three visualizations to show the value distribution of children who received the selected prediction score.
For binary factors, a progress-bar like design is used.
For numeric factors, a box-and-whiskers plot shows the global minimum and maximum values for a feature, and the minimum, first quartile, third quartile and maximum for the selected risk score.
For categorical factors, segmented bars are used to encode a categorical value distribution; hovering over a segment provides more information about the categories and their corresponding percentages. 
}

This \page{} was added as we theorized it may help screeners build trust in the model \TR{} by seeing how it generally performs, and it may clarify the value of the prediction target \CT{} by showing how it relates to a more tangible output of removals from the home.

\add{
\textbf{Feedback.}
Like the \featurecap{} Importance \page{}, screeners expressed concern that the \featurecap{} Distributions \page{} shows too much information for use during active decision-making. However, they said it may be useful for training and finding gaps in provided services.
}

\begin{figure*}[b]
    \centering
    \includegraphics[width=\linewidth]{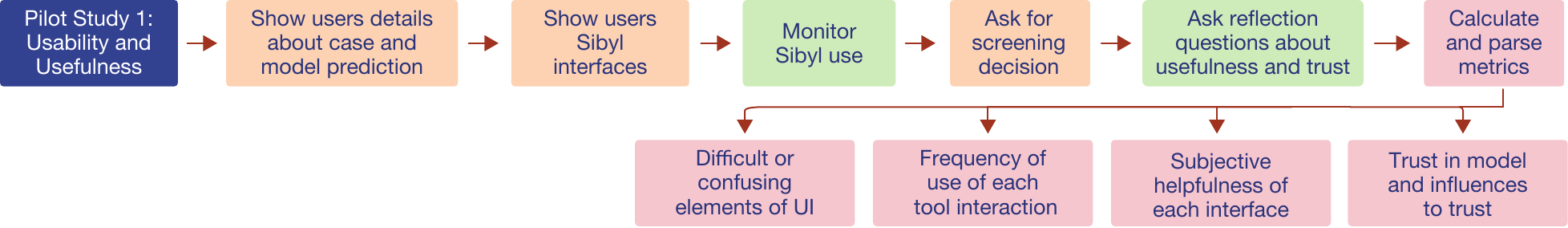}
    \caption[User study procedure.]{The procedure for our formal user study. Our participants were first shown the description of potential abuse from a child welfare referral, as well as the corresponding ML prediction risk score. Next, they were given the opportunity to interact with the \Sibyl{} interfaces. Once they were ready, they were asked to make a screen-in or screen-out decision, and then asked a series of reflection questions.}
    \label{fig:pilot}
\end{figure*}

\section{\add{User Studies}}\label{sec:user_study_1}

To evaluate \Sibyl{}, we ran two formal user studies.

\add{Our first study involved 12 data and/or social scientists. We chose to run a study with non-experts first in order to fix immediate usability problems and iterate on the UI/UX elements of the tool. Although these participants had no prior experience in child welfare screening, it was possible for them to understand the process intuitively enough to use the tool in roughly the same way as experts would.}


In our second study, we engaged 13 collaborating child welfare screeners (experts), 2 of whom completed the task while video conferencing and screen-sharing with us. We discuss the results of both user studies in this section.

\noindent \textbf{Data used}: \add{For privacy reasons, we used only synthetically generated and deidentified data in this section. Data for different factors was generated using the CTGAN synthetic data generation algorithm \cite{xu2019modeling}, in which a generative model is learned from the real data and samples that resemble the real data are drawn from the model.}

Case descriptions were real paragraph-form narratives provided by concerned parties during past referrals. Names were changed by a representative of the county for de-identification. For example:
\begin{quote}
    \textit{``Caller (teacher) says Abby (age 5) came into school with bruise on arm. Caller says Abby often comes in bruised. Abby told teacher she fell off bike. Teacher asked Abby's mother about this and mother started acting aggressive...''}
\end{quote}

\subsection{Study Procedure}

Participants were first shown a short video explaining how to use \Sibyl{}. Next, they were shown 7 case descriptions, accompanied by model predictions and \Sibyl{} \pages{} with simulated data. 

Participants were then asked to make a screen-in/screen-out decision, and to answer some reflection questions. These questions included 1) five-point Likert-scale style questions about how much participants trusted the model and how confident they felt in their decisions, 2) multiple choice questions about which \Sibyl{} interfaces were helpful, and 3) free response questions about trust in the model and general feedback. 

In total, experts completed 73 individual case analyses, and non-experts completed 75. The procedure for this user study is summarized in Figure \ref{fig:pilot}. 

\subsection{Study Results} \label{sec:changes_needed}

\subsubsection{Helpful Interfaces} \label{sec:useful_interfaces}
To address \textbf{RQ2}, we analyzed the self-reported helpfulness of each augmentation \page{}.

The Case-Specific Details \page{} was by a large margin considered the most helpful \page{}, by both experts and non-experts. It was labelled as being helpful by experts in 91.8\% of case analyses, and by non-experts in 90.7\% of case analyses. This was significantly higher than Sandbox (experts: 16.4\%, non-experts: 22.6\%) and \featurecap{} Distributions (experts: 20.5\%, non-experts: 8.0\%). \featurecap{} Importance was never listed as helpful by either group. 

\subsubsection{Reliance on Sibyl}
Unsurprisingly, non-experts were more likely to report listening to the model without considering the added information in \Sibyl{}. One non-expert participant commented, 
\begin{quote}
    ``\textit{No idea what is going on in this case description -- so completely defer to the model here.}''
\end{quote} 
Another non-expert participant commented 
\begin{quote}
    ``\textit{I found the score useful - and used it as a justification for screening out without exploring in detail all the factors.}''
\end{quote} 
Additionally, non-experts reported that they used the model ``a lot'' or ``a great deal'' on 46.4\% of cases, while experts chose these options in 15.6\% of cases. 

\subsubsection{Impact on Trust}

The \Sibyl{} \pages{} were reported to both increase and decrease expert users’ trust in the model for different reasons. \add{To analyse why this might occur, we thematically analyzed the responses to the open question ``What made you trust the model more or less?’’ We gathered 56 responses to this question and divided the answers into two categories, based on the corresponding answer to the 5-point Likert scale question ``How much did you trust the model's prediction for this case?’’ 43 free-responses corresponded to trusting the model ``a great deal,’’ ``a lot,’’ or ``a moderate amount,’’ while 13 corresponded to trusting the model ``a little'' or ``not at all.’’ 
For all but three of these answers, the response corresponded with the degree of trust listed (ie., participants who reported trusting the model provided reasons why they trusted the model \textit{more}). We do not include the three exceptions in our analysis.}

\add{To conduct the thematic analysis, we sorted according to which (if any) specific \Sibyl{} interfaces were referenced, and further identified what specific elements of the interfaces were referenced. For answers that did not reference a specific interface, we identified themes in the responses, such as agreement with the model, references to general information provided by the model, or inaccuracies in the model. With these two coding mechanisms, we sorted the responses into 7 themes that corresponded with increasing trust, and 5 that corresponded with decreasing trust.}

Table \ref{tab:trust} lists these themes. We see that agreeing with the model's score increases trust of the model the most. Beyond this, the Case-Specific Details page was frequently cited as increasing model trust, either due to specific factors listed or more general elements of the page, such as the number of factors. Trust was reduced when there was confusion or inconsistencies in the presented information, or when the model did not consider important factors that participants knew about. 

\subsubsection{Information Presentation} \label{sec:design_changes}
To address \textbf{RQ3}, we categorize and summarize the comments made by users regarding \Sibyl{} design choices, as well as the steps we took to address them.

\begin{enumerate}
    \item \textbf{Too many \features{} shown \quad} The model was originally trained on over 400 \features{}, \add{all of which were presented in the \Sibyl{} interface,} but many of these \features{} have zero or near-zero weight. For example, one participant commented: 
    \begin{quote}
        ``\textit{Too many factors listed. I only want to see the material risk and protective factors.}''
    \end{quote}    
    Our updated version of \Sibyl{} only shows 10 \features{} by default, with an option to show more. 
    
    \item \textbf{Confusion caused by correlated \features{} \quad} The model uses some engineered \features{}, resulting in \features{} that have deterministic relationships. For example, there is a numeric \feature{} called \texttt{AGE OF CHILD}, and then a set of binary \features{} referring to each age group: i.e. \texttt{CHILD IS LESS THAN 1 YEAR OLD}, \texttt{CHILD IS BETWEEN THE AGES OF 1 AND 3}, etc. These \features{} may cause confusion when shown directly to users. In addition to increasing the cognitive load on users without providing additional useful  information, explanations using these \features{} may reveal seemingly contradictory or unusual relationships. For example, an age category may contribute greatly, while the numeric age \feature{} does not. 
    
    Additionally, having these correlated \features{} causes confusion on the sandbox page, as it is possible to change one \feature{} without changing all of the other deterministically-correlated \features{} in its set. One participant commented, 
    \begin{quote}
        \textit{``I'm not sure, in the sandbox, if I change one feature, other features will be changed automatically.'' }
    \end{quote}
    
    To solve these problems, we combined the correlated \features{} in the \Sibyl{} interface, forming categorical \features{} out of binary one-hot encoded \features{}, and summing the additive contributions.
    
    \item \textbf{Confusion caused by Boolean terminology \quad} One source of confusion was the method of displaying Boolean \features{}. In our original design, we displayed the description of the \feature{}, with a value of \texttt{True} or \texttt{False}. This is the most accurate way of representing the model's logic, but it is not the most intuitive way for our end-users. One participant said, 
    \begin{quote}
        \textit{``The `true' and `false' is hard to interpret... Would rather have a positive statement (e.g., no perpetrator named)''}
    \end{quote}
    
    Therefore, our final version of \Sibyl{} instead states only true statements about the child --- including by negating descriptions of false \features{}. For example, the \feature{} \texttt{CHILD HAS SIBLINGS} with a value of \texttt{False} will be displayed as \texttt{CHILD DOES NOT HAVE SIBLINGS}.
\end{enumerate}

\begin{table*}[t]
\caption{Summary of answers to the question ``what made you trust the model more/less''. The top section lists reasons for having ``a great deal'', ``a lot'' or ``a moderate amount'' of trust in the model. The bottom section lists reasons for having ``a little'' or ``not at all'' trust in the model. The first column lists the general themes we found in the answers. The second column lists the number of answers that fell within each theme. The final column lists selected answers for each theme.}
\footnotesize
\begin{tabular}{{p{0.4\textwidth}p{0.06\textwidth}p{0.44\textwidth}}}
\toprule
\multicolumn{3}{l}{Factors that increased trust} \\
\midrule
Category                                                                                                                      & \begin{tabular}[c]{@{}l@{}}\# of \\ answers\end{tabular} & Sample answers                                                                                                                                                                                                                 \\ \midrule
General comments about being shown protective and risk factors (Details page)      & 8                                                        & ``\textit{The details and the risk and protective factors and the contribution they have}'' ``\textit{Info in the details and risk factors}''                                                            \\
Specific factors listed as risk or protective (Details page)                    & 8                                                        & ``\textit{The past number of child welfare involvements (listed in the features listed)}'', ``\textit{The risk factors involved, especially prior placements, benefits, and current CPS involvement}'' \\
The number of factors listed (Details page)                                                                                                  & 5                                                        & ``\textit{Very few risk factors}'', ``\textit{The lack of protective factors}''                                                                                                                                                                      \\
The score agreed with screener intuition                                                                                      & 10                                                       & ``\textit{...not residing with the alleged perpetrators which I would assume would reduce the risk score}'',  ``\textit{Model prediction makes sense}''                             \\
The explanation agreed with screener intuition                                                                                & 3                                                        & ``\textit{Risk factors made sense for the model prediction number}''                                                                                                                                                                      \\
General comments about sandbox page                                                                                           & 1                                                        & ``\textit{Details under sandbox of why the risk level was so high}''                                                                                                                                                                      \\
General comments about the explanation providing more information or understanding & 6                                                        & ``\textit{Allowed for more understanding}'', ``\textit{History clarification}''                                                                                                                                                                      \\ \bottomrule

\toprule
\multicolumn{3}{l}{Factors that decreased trust} \\
\midrule
Category & \begin{tabular}[c]{@{}l@{}}\# of \\ answers\end{tabular} & Sample answers                                                                                               \\ \midrule
A specific, key factor was not considered by the model & 3                                                       & ``\textit{This is a case for law enforcement, not CPS}'', ``\textit{...it may have been handled during the open case}''           \\
The importance weighting of factors was off (Details page)            & 2                                                       & ``\textit{Young, vulnerable children being left alone is still cause for concern, despite past involvement}''           \\ 
The score disagreed with screener intuition            & 3                                                       & ``\textit{Seems high, no health history, no real proof of any drug use, no proof child is at any risk of abuse}'' \\
There was some confusion about information presented   & 2                                                       & ``\textit{There were discrepancies between the info in the referral and the info provided by the tool}''                \\
The screener wanted more information                   & 2                                                       & ``\textit{I would want to see other referrals for the family}''  
 \\ \bottomrule
\end{tabular}
\label{tab:trust}
\end{table*}

\section{Discussion and Limitations} \label{sec:discussion}
\add{In this section, we summarize general lessons learned regarding how augmentation tools can improve the usability of ML models in the child welfare domain. We speculate that these lessons may generalize to other domains with similar context factors --- i.e. those where users with high domain expertise and low technical expertise make high-impact/high-risk decisions with the help of ML models. }



\subsection{Helpful Augmentation Tools}
\add {In Section \ref{sec:useful_interfaces}, we noted that our user study participants did not frequently list our About Model or Sandbox interfaces as helpful, despite our hypothesis that these interfaces could address some usability challenges as noted in Table \ref{tab:interfaces}. While further user studies will be required to identify exactly why more interfaces were not used, a few potential reasons come to mind. First, screeners are used to making decisions fairly quickly, and may not have enough time to parse and consider case details while investigating more than one interface. Second, the About Model interfaces may not be very relevant to screeners in the midst of a specific case, and may be more helpful if presented to screeners prior to screening. Third, the experimentation functionality of the Sandbox page is only useful if screeners have a specific what-if question in mind, while the “flipped feature” aspect may be somewhat redundant – it shows similar information to the Case-Specific Details page (the relative importance of each factor), but it a way that appears to be less intuitive to users. }

\subsection{Accuracy versus Fidelity}
Robnik-Sikonja and Bohanec \cite{robnik-sikonjaPerturbationBasedExplanationsPrediction2018} define the \textit{accuracy} of an explanation as how well it generalizes to other unseen examples (i.e., how accurately these rules predict what happens in the real world), and \textit{fidelity} as how well an explanation describes the model itself.

Our users were mostly interested in getting \textit{accurate}  explanations that provided information about \textit{the case at hand}. As evidenced by all three findings about the \page{} design (Section \ref{sec:design_changes}), users wanted to receive information about the model in a language and format that mirrored their own, not the format used by the model itself. This is also evidenced by design requests like using the terms \textit{risk and protective factors}, rather than the more ML-centric terms \textit{negative and positive~features}.

\subsection{The Importance of Interpretable Factors}
Simple models, such as regression, are often cited as being inherently interpretable \cite{rudinStopExplainingBlack2019}. However, this case study suggested that even simple models may cause confusion in users, and lead to challenges when attempting to explain model predictions for decision making. 

Instead, our work found that, for the purposes of making models usable for end-users, the interpretability of the model \features{} may be most important. In our study, the screeners were often confused when explanations used \features{} that did not have clear implications on risk. For example, in our user study, one participant said 
\begin{quote}
    \textit{``... 2 parents have missing date-of-birth is shown as a significant blue bar which I can't imagine is protective.''}
\end{quote} 
Additionally, as discussed in Section \ref{sec:design_changes}, one-hot-encoded \features{} were not interpretable, and many of the reasons screeners trusted the model more or less (Table \ref{tab:trust}) related to the specific \features{}. 

\subsection{Non-Applicable Usability Challenges}

One interesting finding of our work was the \problems{} we did not see evidence of in this case. 

For example, one possible use of explanations is to give humans the ability to actively correct errors in a model's logic. We did not see evidence of this behavior from our users, however. There are several possible reasons for this. First, our users are making decisions in a very limited time, and do not have additional time to review the model's quality. Second, our users are thoroughly analyzing every case on their own and were only using the model as an extra flag. Finally, the users already have some discomfort about the model, likely due to the high associated risk. As a result, our users tended to discount the model altogether if they did not believe it was correct about a particular case.

Additionally, users expressed almost no interest in learning about the model itself through explanations. A common explanation need addressed by the literature \cite{liptonMythosModelInterpretability2016} is to understand how the model works (model transparency), possibly for debugging. In our case study, however, only once (see Section \ref{sec:interviews}, Item \ref{item:steps}) did any screener express interest in understanding the details of how the model worked under the hood --- and even then, they were mostly looking for a broad overview. 

\subsection{Cognitive Biases} \label{sec:cog_bias}
Wang et. al. \cite{wangDesigningTheoryDrivenUserCentric2019} introduced a list of the cognitive biases explanations can help address. Our experience with child welfare screeners additionally suggested some of these cognitive biases could be \textit{encouraged} by the explanations and other forms of further information. For example: 

\begin{enumerate}
    \item \textbf{Representativeness Bias \cite{wangDesigningTheoryDrivenUserCentric2019}}: Case-based explanations that offer similar examples to the case at hand (such as our Similar Cases page) risk encouraging users to make decisions based on similarities to another case.
    \item \textbf{Causation vs Correlation}: Counterfactual-based explanations, which consider how the model prediction would change under different circumstances, made participants more likely to interpret the explanations as containing information about the causal structure of the world.
    \item \textbf{Availability Bias \cite{wangDesigningTheoryDrivenUserCentric2019}}: A \feature{}-contribution explanation that is sorted in ascending order (and therefore lists negative contributions first) may result in different decisions than one that is sorted in descending order (and therefore lists positive contributions first) due to availability bias, which causes humans to put too much importance on recent or memorable events or information.
\end{enumerate}

Further work and user studies may better reveal the extent to which these biases are caused or exacerbated by ML augmentation tools.

\subsection{Limitations}
Our work has some limitations that may be addressed in future work.

\add{Our analysis focused on the qualitative comments and self-reported confidence measures provided by participants, which we used to measure the usability and perceived usefulness of the different explanation interfaces. Because the data used in our formal user study was synthetic, and therefore not associated with real-world outcomes, we did not measure the efficacy of the screening decisions made by these users. Arguably, it would be possible to quantitatively measure certain effects of \Sibyl{} -- such as changes in the quality of decisions made, or differences in results depending on how the platform was deployed -- or even to use \Sibyl{} to identify biases in human decision-making. However, this was beyond the scope of this paper, which sought to explicitly augment human decision-making, rather than potentially cross the line into partially automating it.}

\add{The literature on ML usability would benefit from a complete, methodological investigation into the ML usability challenges present across domains, to extend our sample subset introduced in Section 1.}

\section{Conclusion and Future Work}

In this work, we identified the ML \problems{} associated with the domain of child welfare screening. We found one promising tool (\feature{} contributions, on the Case-Specific Details \page{}) for mitigating many of these challenges, and pinpointed important design decisions that must be made to maximize the usefulness of this tool. \add{Future work should empirically investigate the effect this tool has on decision-making, quantitatively measure how well it mitigates existing \problems{}, and explore innovative visualization designs to better solve the remaining challenges.}


\acknowledgments{
We would like to thank Rhema Vaithianathan, Diana Benavides Prado, Megh Mayur, Athena Ning, and Larissa Lorimer for their guidance throughout the process of applying machine learning explainability to child welfare, for connecting us to the child welfare domain experts, and for providing us with their trained model and dataset. We thank the child welfare experts at Larimer County Department of Human Services for their invaluable insights and feedback on the tool. We thank Iulia Ionescu, Sergiu Ojoc, Ionut Radu, and Ionut Margarint for their work on developing the Sibyl application. We thank Arash Akhgari for his work on our diagrams and visualizations. We thank Michaela Henry for support in project management and insights. We thank the participants of both user studies for their time and feedback. Finally, we would like to thank our anonymous reviewers for their comments and suggestions. 
This work is supported in part by NSF Award 1761812.}

\bibliographystyle{abbrv-doi}

\bibliography{references}

\appendix
\newpage

\section{Appendix: Mitigating Tool Definitions} \label{app:defs}
Here, we define the types of mitigating tools discussed in this paper and Table \ref{tab:usability_concerns}.

\textbf{Global Explanation:} An explanation of a model's general logic, achieved through methods such as quantifying the overall importance of features or visualizing the model boundary \cite{doshi-velezRigorousScienceInterpretable2017}.

\textbf{Local Explanation:} An explanation as to why a model made an individual prediction, achieved through methods such as quantifying how much each feature contributed to this particular prediction \cite{doshi-velezRigorousScienceInterpretable2017}.

\textbf{Cost-Benefit Analysis:} A measurement of the expected total reward from taking an action, defined by the expected benefits minus the costs \cite{kentonHowCostBenefitAnalysis2021}. In the case of machine learning, this would involve providing information about the expected results of a prediction alongside the prediction itself. 

\textbf{ML Fairness Metrics:} Mathematical approaches to measuring the level of bias present in models \cite{googleMachineLearningGlossary2020}

\section{Appendix: Design Mockup Reviews} \label{app:mockups}
\begin{figure*}
    \centering
    \includegraphics[width=\textwidth]{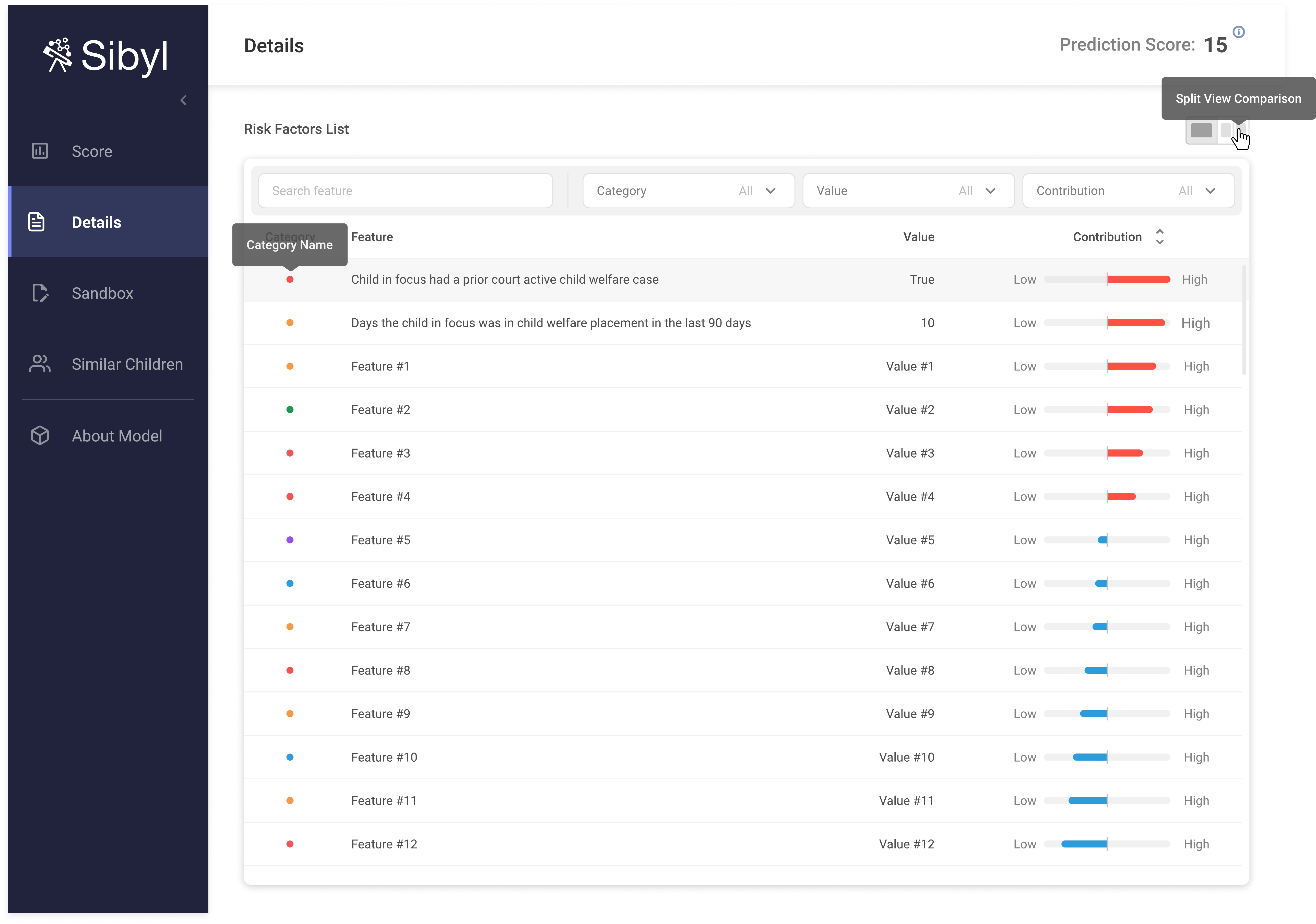}
    \caption{First draft of the \textit{Details} page mockup. Initial interview feedback suggested that this was the most helpful page, so it was kept as the first and default option. Additionally, screeners said that in their usual workflow, they would list ``Risk'' and ``Protective'' factors side by side. To mirror this, the updated version of this page has a split-view toggle that shows the negative- and positive-contribution features in two side-by-side tables.}
    \label{fig:first_details}
\end{figure*}

\begin{figure*}
    \centering
    \includegraphics[width=\textwidth]{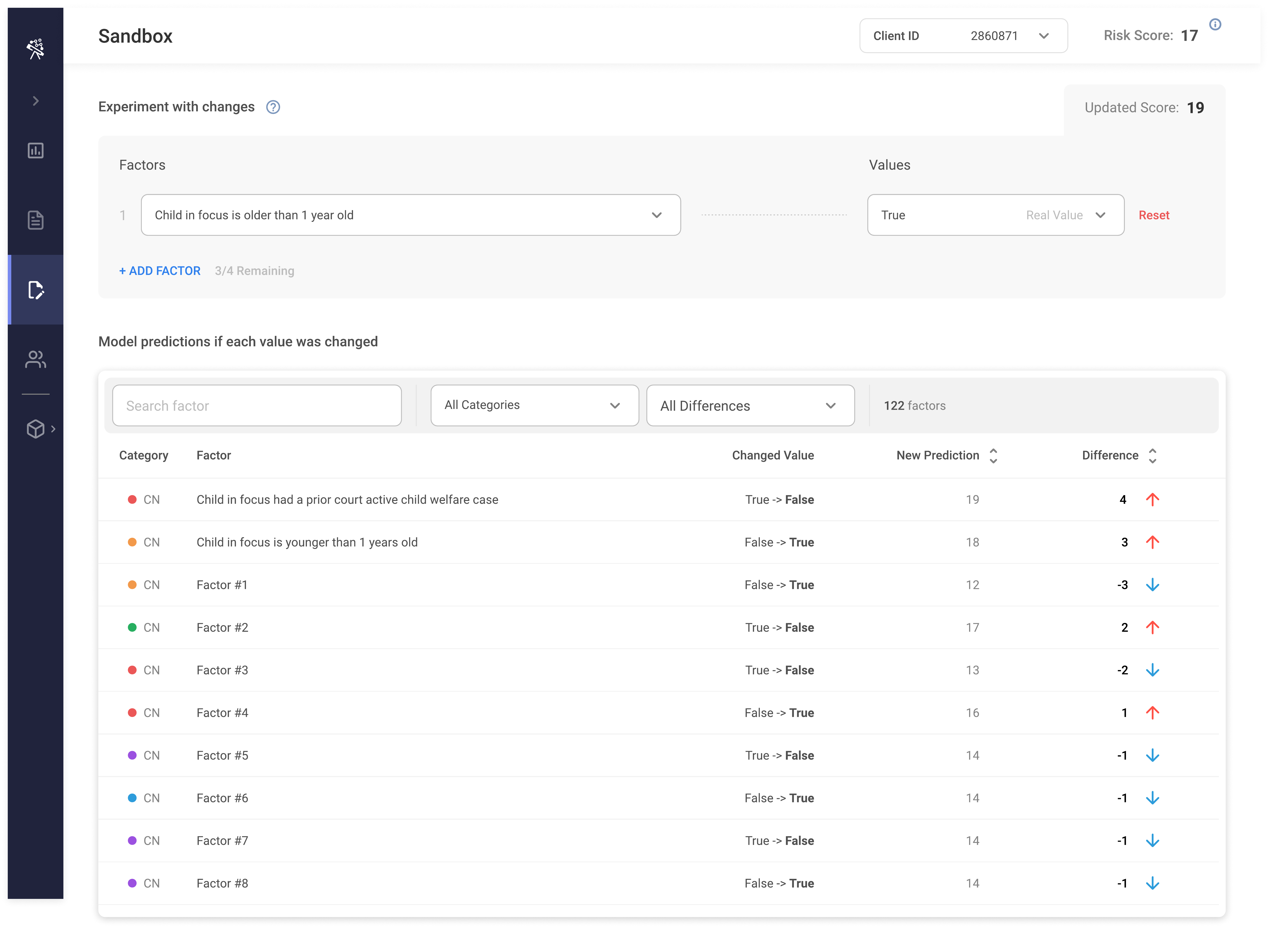}
    \caption{First draft of the \textit{Sandbox} explanation page. Feedback confirmed our concern that this view has a risk of being misconstrued as suggesting action or reflected real-world causal structures. However, screeners also said that they saw value in this screen as a supervision tool, used to review model and human decisions rather than actively during the decision-making process. \newline Screeners also pointed out that this screen appeared to provide similar information to the Details page, and therefore may not be needed.}
    \label{fig:first_sandbox}
\end{figure*}

\begin{figure*}
    \centering
    \includegraphics[width=\textwidth]{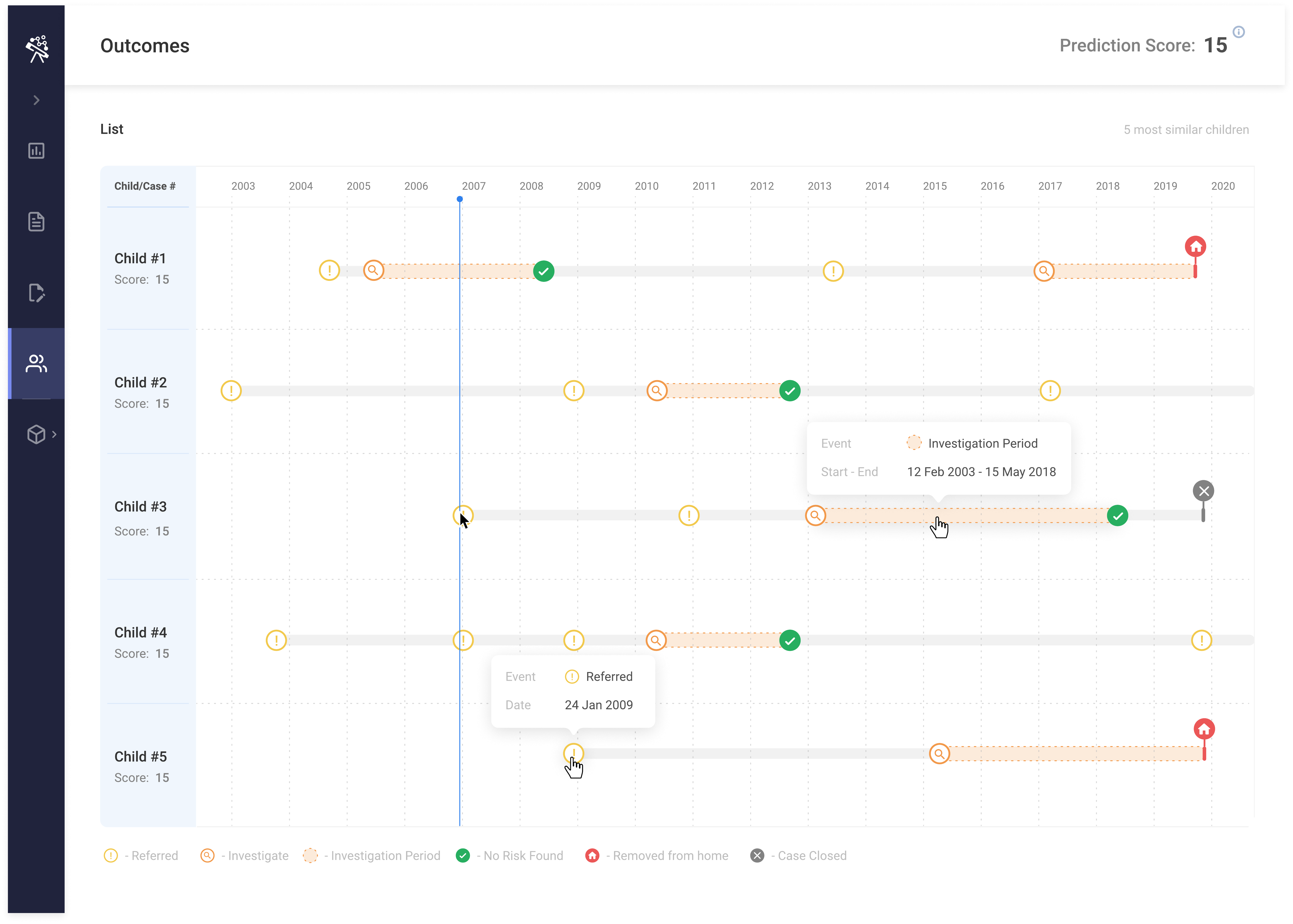}
    \caption{The first draft of the \textit{Similar Children} explanation page.  Screeners were concerned that this page may cause poor decision making, as making decisions on a case based on past cases that seem similar is discouraged. Screeners reported that this kind of thinking can lead to bias or self-fulfilling prophecies. Therefore, it was decided that this screen would not be used at decision time. \newline County officials pointed out that this kind of explanation has been used in the past to justify unusual model predictions during review. For this use case, however, a more detailed view of what exactly happened in the case would be more useful. Therefore, we modified this view to feature detailed timelines of each similar child instead, and decided to test it as a supervision tool instead.}
    \label{fig:first_similar}
\end{figure*}

\begin{figure*}
    \includegraphics[width=\textwidth]{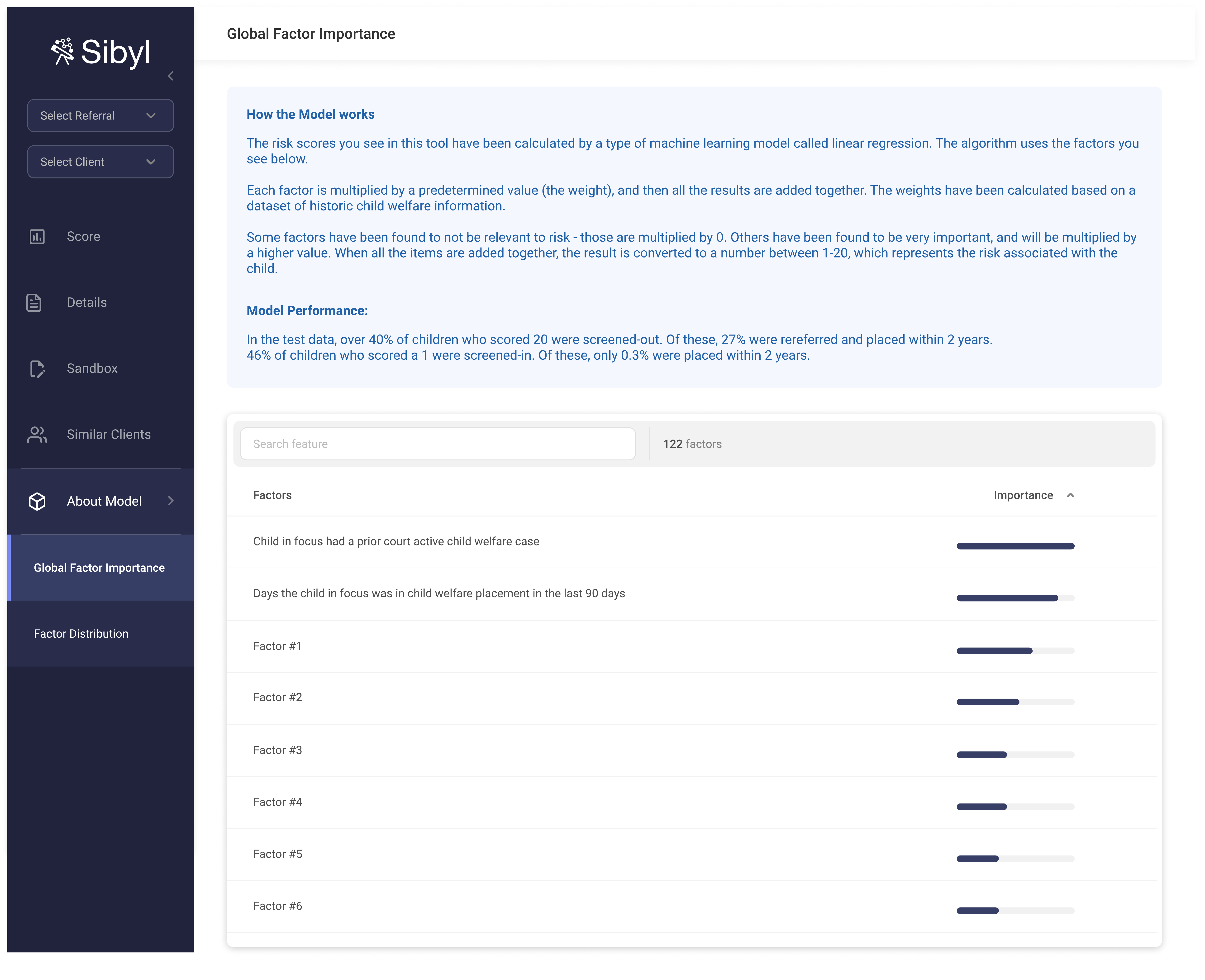}
    \caption{The first draft of the \textit{\featurecap{} Importance} page. Screeners said that this screen seemed intuitive, but may provide too much information to be useful during active decision-making, and may be better for training and education. }
    \label{fig:first_importance}
\end{figure*}

\begin{figure*}
    \includegraphics[width=\textwidth]{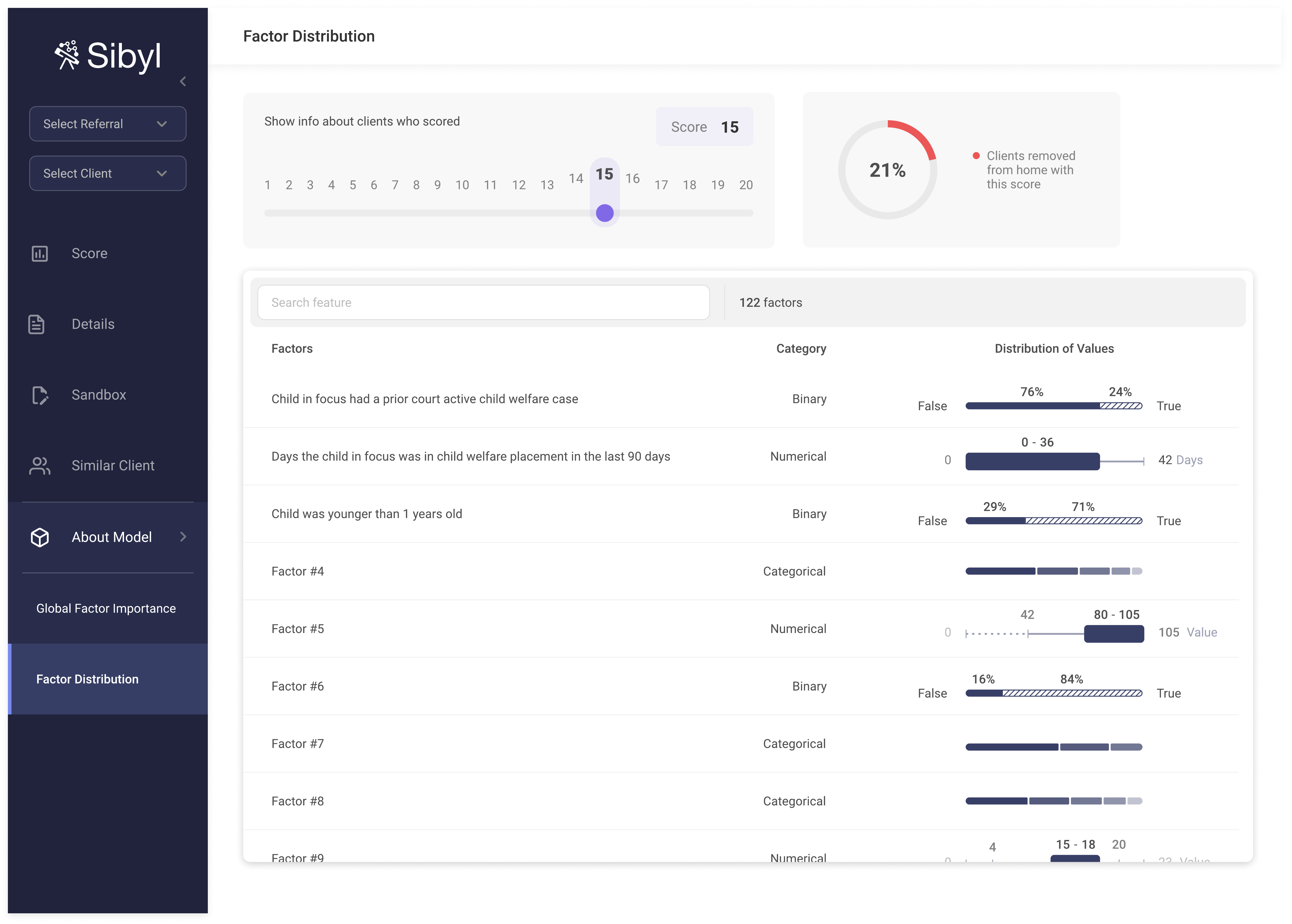}
     \caption{The first draft of the \textit{\featurecap{} Distributions} page. As with the Feature Importance page, screeners expressed concerns that this page shows too much information for use during active decision-making. However, they said that it may be helpful for use in training, and for finding gaps in provided services.}
    \label{fig:first_distributions}
\end{figure*}

Figures \ref{fig:first_details} - \ref{fig:first_distributions} show the original design mockups that were presented to child welfare screeners, and describe the feedback and changes that were made as a result of this interview.

\section{Appendix: User Study Questions Asked} \label{sec:questions_asked}

Table \ref{tab:questions_asked} contains the complete list of questions we asked during our user study.

\begin{table*}[htb]
\centering
\begin{tabular}{{p{0.15\textwidth}p{0.2\textwidth}p{0.55\textwidth}}}
\toprule                                                    
\textbf{When?}                   & \textbf{Response Type}   & \textbf{Question}                                                 \\ \midrule
Beginning                        & Multiple choice          & What is your experience with child welfare screening?             \\ \midrule
\multirow{7}{*}{After each case} & Multiple choice          & Would you choose to screen-in or screen-out?                      \\
                                 & 5-point Likert scale     & How confident are you in your decision?                           \\
                                 & 5-point Likert scale     & How much did the prediction score impact your decision?           \\
                                 & 5-point Likert scale     & How much did you trust the model’s prediction for this case?      \\
                                 & Free response            & What caused you to trust the model more or less?                  \\
                                 & Multiple-multiple choice & What explanations, if any, did you find helpful?                  \\
                                 & Free response            & Any other comments?                                               \\ \midrule
\multirow{4}{*}{End}             & 5-point Likert scale     & How helpful did you find the model's predictions overall?         \\
                                 & 5-point Likert scale     & How helpful did you find the Sibyl tool's explanations?           \\
                                 & Multiple-multiple choice & Which explanation did you find most helpful overall?              \\
                                 & Free response            & Were there any feature categories you found more or less helpful? \\ \bottomrule
\end{tabular}
\caption{Questions asked in the UI/UX and Outcome Evaluation user studies.}
\label{tab:questions_asked}
\end{table*}

\end{document}